# Comparative study
# of different functionalized graphene-nanoplatelet aqueous nanofluids for solar energy applications

**Javier P. Vallejo[1], Luca Mercatelli[2], Maria Raffaella Martina[2], Daniele Di Rosa[2,3], Aldo Dell'Oro[4], Luis Lugo[1], Elisa Sani[2]\***

[1]Departamento de Física Aplicada, Facultade de Ciencias, Universidade de Vigo, E-36310 Vigo, Spain
[2]CNR-INO National Institute of Optics, Largo E. Fermi, 6, I-50125 Firenze, Italy
[3]University of Pisa, Department of Energy, Systems, Territory and Constructions Engineering (D.E.S.T.eC.), Largo L. Lazzarino, I-56122 Pisa, Italy
[4]INAF, Osservatorio Astrofisico di Arcetri, Largo E. Fermi 5, I-50125 Firenze, Italy
*corresponding author: elisa.sani@ino.cnr.it

**Abstract**
The optical properties of nanofluids are peculiar and interesting for a variety of applications. Among them, the high light extinction coefficient of nanofluids can be useful in linear parabolic concentrating solar systems, while their properties under high light irradiation intensities can be exploited for direct solar steam generation. The optical characterization of colloids, including the study of non-linear optical properties, is thus a needed step to design the use of such novel materials for solar energy exploitation. In this work, we analysed two different types of nanofluids, consisting of polycarboxylate chemically modified graphene nanoplatelets (P-GnP) and sulfonic acid-functionalized graphene nanoplatelets (S-GnP) dispersed in water, at three concentrations from 0.005 wt% to 0.05 wt%. Moderately stable nanofluids were achieved with favourable light extinction properties, as well as a non-linear optical behaviour under high input solar intensities.

**Keywords:** direct absorption solar collectors; concentrating solar power; solar steam generation; nanofluids; graphene nanoplatelets; optical properties

## 1. Introduction

Nanofluids (i.e. suspensions of nanometer-sized solids in a base fluid) are receiving a huge development in last years and are taken into account for a large variety of applications. Among the fundamental characteristics of nanofluids, the knowledge of optical properties is needed for all applications requiring the interaction of the nanomaterial itself with optical radiation. This field results, in fact, really huge: from nanobiology/nanomedicine [1-4] to different types of optical sensors [4-7], from the development of innovative light sources and optical components [3, 8] to nanomotors [9, 10], just to mention few of them. Energy is one of the most appealing application fields of nanofluids, and, in particular, dealing with optical properties, the solar energy exploitation. [11, 12].

As confirmed by the impressive growth of studies on this topic [13-16], dark nanofluids are considered promising for the direct-absorption solar collector (DASC) scheme. DASCs offer several advantages with respect to conventional surface-absorber architectures, because volumetric sunlight absorption and heat exchange functions are carried out by the same element, namely the dark fluid, arising in an optimized thermal transfer and in reduced thermal re-radiation losses. The DASC system was originally proposed by Minardi and Chuang [17] in the mid-1970s to potentially enhance the system efficiency by absorbing the energy within the fluid volume [18]. In particular, this was typically done using India Ink, a black fluid composed by a suspension of micron-sized carbonaceous particles prone to sedimentation and clogging of the pumps [13]. Moreover, black inks usually contain organic and inorganic chromophores that show serious drawbacks because of their light-induced and thermal degradation at the high operating temperatures. [18, 19]. Nanoparticles overcome these critical issues and potentially increase the efficiency of DASCs. A large variety of nanoparticles and base fluids has been investigated in the literature. Among them, the family of carbon-based nanostructures has emerged as particularly promising. From carbon black [20] to graphite [21], as well as single, multi-wall and functionalized nanotubes [22-24], carbon nanohorns [20, 25-27], many nanofluids containing carbon nanoparticles have been studied for solar energy applications. Graphene is one of most intriguing carbon allotropes [28, 29]. Graphene nanoplatelets or nanosheets consist of small flakes of several-layer staked graphene that partially inherit the good properties of graphene with much lower production costs. Graphene nanoparticles are naturally hydrophobic. They can be stably suspended in water after chemical modifications of their surface, e.g. by oxidation or by different types of functionalization [30, 31]

A second promising field of application of nanofluids in renewable energies is direct solar steam generation. Steam production is of crucial importance for many applications including electricity generation, energy storage, biomass processing, water desalination and sterilization [32-35]. Green and renewable steam production by solar energy is thus an important topic of research [36], as an example to bring technologies essential for life in off-grid areas and resource-poor locations. Typical solar steam production systems are currently based on solar trough or solar tower architectures with a surface or cavity solar absorber [37-39]. They work heating a bulk fluid to its boiling temperature under high optical concentration. The steam generation efficiency is strongly connected to the surface temperature and thermal radiation properties of the absorber. However, as bulk steam production needs high temperatures, these conventional systems also suffer from high heat losses and low efficiency. To overcome this drawback, a possibility is looking for solutions able to produce steam without requiring heating the whole liquid volume to the boiling point. Different authors reported on the efficient solar steam generation by different nanoparticles, such as gold [40-42], silicon [43] and carbon nanostructures [44, 45].

For the solar steam production, the nanofluid must be subjected to intense light irradiation, as it happens under high solar concentration regimes. It is important to notice that when materials are maintained in such conditions, their response could be nonlinearly dependent on the input light intensity. In this regime, basic optical properties like transmittance undergo a change, and the values obtained under low input intensities are no more applicable. For this reason, a dedicated study must to be carried out, to assess intensity-dependent optical properties as well.

The obtaining of homogeneous dispersions of the nanoadditives in the base fluids is still nowadays a methodological challenge and this issue usually must be confronted [46]. The strong van der Waals interactions between nanoparticles favour the creation of aggregates that tend to easily settle because of their higher weights, leading to variations in the dispersion physical properties [47]. Different physical or chemical methods are usually employed to obtain stable samples as ultrasonication, addition of surfactants, pH-modification or the surface modification of the particles [46].

In this work, we report on the comparative characterization of linear and nonlinear optical properties of graphene nanoplatelets showing two different functionalization types and suspended in water. These suspensions were characterized through a stability analysis based on zeta potential and dynamic light scattering (DLS) measurements. Then, we focused on the DASC application, assessing the spectral extinction coefficient of aqueous nanofluids as a function of the nanoparticle concentration. Finally, we investigated the nanofluid response to high intensity laser radiation, to evaluate the potential of these nanoadditives for direct vapor generation and other nonlinear optics applications.

## 2. Experimental

### 2.1. Materials and sample preparation

Polycarboxylate chemically modified graphene nanoplatelets, P-GnP, and Sulfonic acid-functionalized graphene nanoplatelets, S-GnP, were supplied by NanoInnova Technologies S.L. Madrid, Spain). Milli Q-grade water, W, was produced by a Milli-Q 185 Plus system (Millipore Ltd., Watford, UK) with a resistivity of 18.2 MΩ·cm.

For each nanopowder type, we prepared three aqueous suspensions, with nanoadditive concentrations of 0.005 wt%, 0.025 wt% and 0.05 wt%. The proper amounts of nanopowder and base fluid were weighted in a Mettler AE-200 balance (Mettler Toledo, Greifensee, Switzerland), with a 0.1 mg uncertainty. After mixing and stirring, samples where submitted to ultra-sonication of 20 kHz frequency during 120 min by means of an Ultrasounds ultrasonic bath (JP Selecta S.A., Barcelona, Spain).

### 2.2. Nanopowder characterization and stability

The size and morphology of the nanoadditives were characterized by transmission electron microscopy (TEM) over the employed nanopowders. TEM analyses were performed through a JEM-2010F field emission electron microscope (JEOL, Tokyo, Japan), working at an operating accelerator voltage of 200 kV. TEM images were obtained over drops of dispersions of each nanopowder in analytical grade ethanol, placed over copper supports and previously dried at room temperature.

The stability of a dispersions can be analysed by UV–VIS spectroscopy, zeta potential analysis, light scattering or electron microscopy, among other methods [46, 47]. In this work, zeta potential and dynamic light scattering analyses were carried out to characterize the stability of the studied samples by means of a Zetasizer Nano ZS (Malvern Instruments Ltd, Malvern, United Kingdom). The zeta potential as a function of temperature was determined for the 0.025 wt% nanofluids in the temperature range from 283.15 to 333.15 K. The evolution of the apparent size of the nanoadditives as a function of the time after preparation for the 0.025 wt% nanofluids was also studied at 298.15 K and using a scattering angle of 173°. Samples in static conditions from the preparation (hereinafter "static" samples) and samples mechanically stirred previously to each measurement (hereinafter "shaken" samples) were analysed. The stirring process applied to shaken samples previous to each apparent size measurement was carried out at 2000 rpm during 1 min by means of a ZX3 Advanced Vortex Mixer (VELP Scientifica SRL, Usmate Velate, Italy). Each value of zeta potential and apparent size presented in this work was obtained as the average between three experimental tests.

## *2.3. Optical characterization*

The optical properties of nanofluids were investigated both in linear conditions, obtaining the spectral extinction coefficient, and under high intensity laser irradiation, assessing optical limiting properties. Spectral optical transmittance in the linearity regime was measured using a double-beam UV-VIS spectrophotometer (PerkinElmer Lambda900) using a variable length cell (50 - 500 μm) and the method described elsewhere [48, 49]. High intensity experiments have been carried out using a pulsed nanosecond Nd:YAG laser as light source (Quantel Q-smart 850, delivering 6 ns pulses at 1064, 532 and 355 nm wavelength, with 10 Hz repetition rate). The three laser emission wavelengths were spatially separated by proper optical elements and focused on the sample by a lens of 300 mm focal length. The sample was held in a quartz cuvette with 10 mm path length, put in a defocused position to avoid cuvette damage. The beam exiting the sample was collected by a couple of lenses of focal lengths 40 and 100 mm and focused on a pyroelectric detector (Ophir PE25C, Ophir PE9C). The energy incident on the cuvette was varied and measured using a pyroelectric detector (Ophir PE50BE, Ophir PE25C, Ophir PE9C). The laser experimental setup is sketched in Figure 1a and the real image is shown in Figure 1b. It is worth to notice that our setup is different, and much simpler than the conventional Z-scan technique. However its use in combination with semi-empirical models allows to identify, in most cases, the active nonlinear process by a simple fitting procedure (see below).

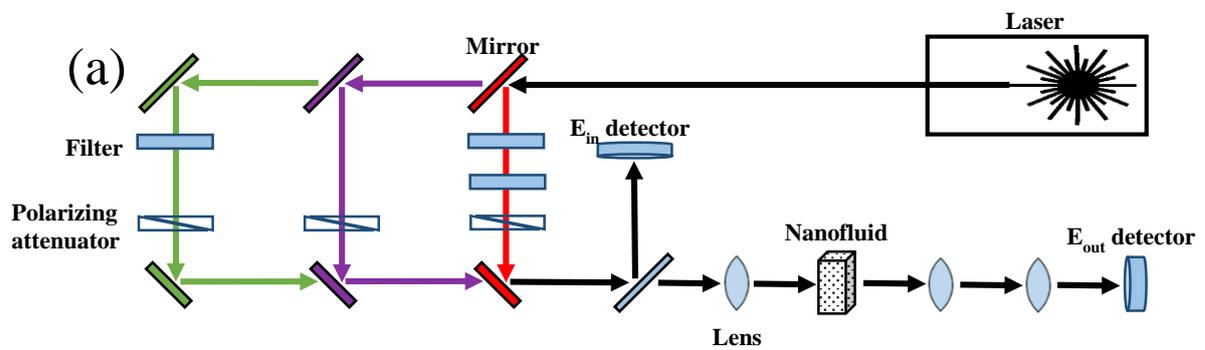

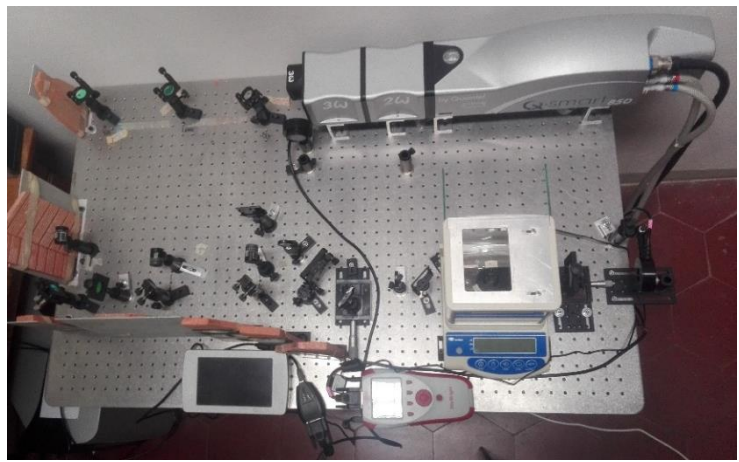

***Figure 1***: *Scheme (a) and real image (b) of the laser experiment setup.*

## 3. Results and discussion

### 3.1 Nanopowder characterization and stability

TEM analysis results permit to settle the nanoplatelet structure of both nanoadditives, that is set of stacked graphene layers, as it can be seen in Figure 2. S-GnP particles are generally larger than P-GnP particles. As an example, Fig. 2a shows a S-GnP of around ~ 380 nm × ~ 590 nm and Fig. 2b shows a P-GnP of around ~ 100 nm × ~ 210 nm.

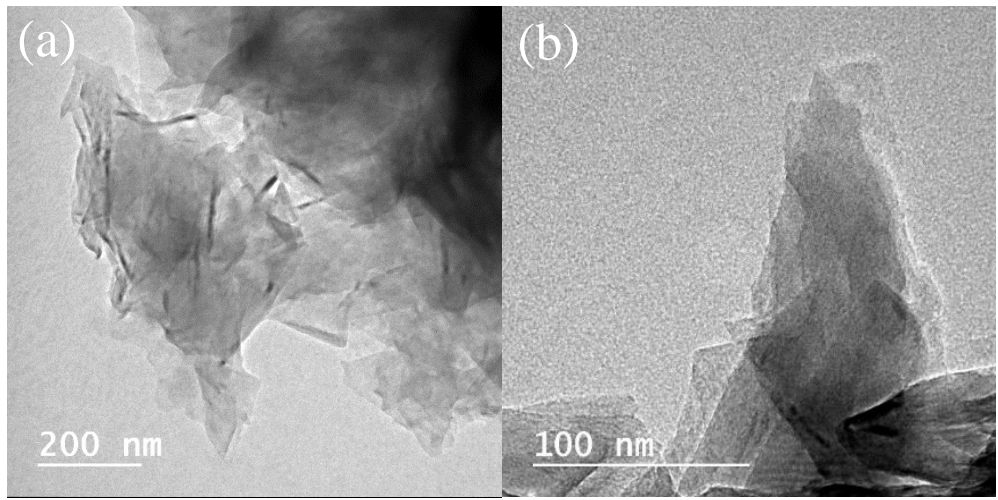

**Figure 2**: *TEM images of the dry employed nanopowders. (a) S-GnP nanopowder and (b) P-GnP nanopowder.*

Previous works showed other structural information of these particles. Atomic force microscopy analyses stated heights per graphene layer of 3 to 12 nm for S-GnP [50] and 2 to 18 nm for P-GnP [51], so there are not significant differences in this dimension. Regarding chemical composition, Energy-Dispersive X-ray Spectroscopy analyses showed the presence of carbon (C), oxygen (O) and sulfur (S) for S-GnP [52] and carbon (C), oxygen (O) and potassium (K) for P-GnP [53].

As for the zeta potential characterization, it should be reminded that the higher are the electrostatic repulsion forces among the nanoadditives, the higher the value of the zeta potential. Usually, absolute zeta potential values over 30 mV are considered as a symptom of nanofluid stability [46, 47, 54, 55], the higher the value, the greater the stability. (Table 1).

| Z potential absolute value (mV) | Stability |
|:---:|:---:|
| 0 | No stability |
| 15 | Possible some stability but settling |
| 30 | Moderate stability |
| 45 | Good stability, possible settling |
| 60 | Very good stability, little settling likely |

**Table 1**. *Relationships between zeta potential and stability [47].*

Figure 3 shows the results of the analysis of the zeta potential as a function of temperature for the 0.025 wt% samples, corresponding to the intermediate concentration among the three analysed. Error bars correspond to the standard deviation of the experimental data. As it can be observed, quasi-constant values were obtained for each nanofluid type, indicator of non-deterioration of samples at the studied temperatures. The mean values obtained were 43.8 and -53.3 mV for the P-GnP and S-GnP nanofluids, respectively. Both values are over the threshold of 30 mV of absolute value, synonym of moderate stability. Nevertheless, the value obtained for the S-GnP/water nanofluid is higher, even over 45 mV of absolute value (good stability, according with Table 1), indicating a potentially better behaviour than the P-GnP/water nanofluid.

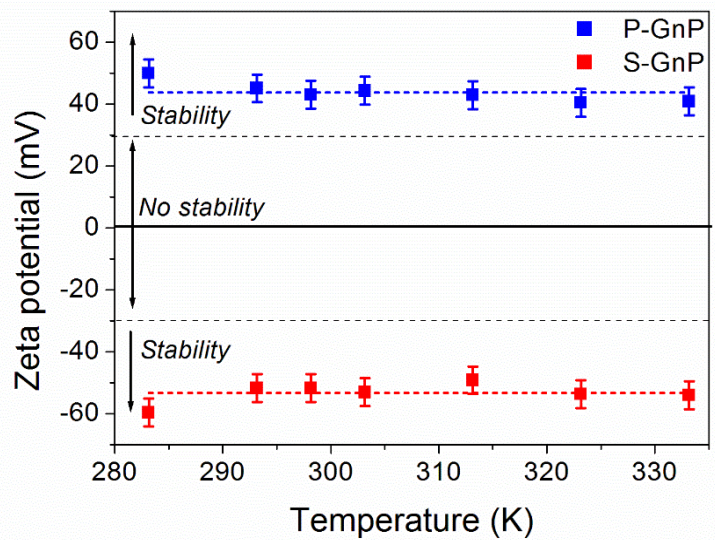

***Figure 3***: *Zeta potential dependence on temperature for the 0.025 wt% samples (red, S-GnP/water nanofluid; blue, P-GnP/water nanofluid).*

Figure 4 shows the size distribution by intensity for the 0.025 wt% S-GnP/water nanofluid and the 0.025 wt% P-GnP/water nanofluid just after preparation and one week after. It should be noted that dynamic light scattering measurements assume that nanoparticles are spherical while the employed nanoadditives are nanoplatelets, so the reported size values should be considered as "apparent" sizes. As it can be observed, in both cases the intensity peak is obtained for practically the same apparent size for the samples measured just after preparation and the shaken samples 7 days later. Nevertheless, the static samples 7 days after preparation showed intensity peaks at lower apparent sizes, indication of the progressive sedimentation of larger nanoparticles. The decrease is comparatively lower for the P-GnP nanofluid, in accordance with zeta potential results.

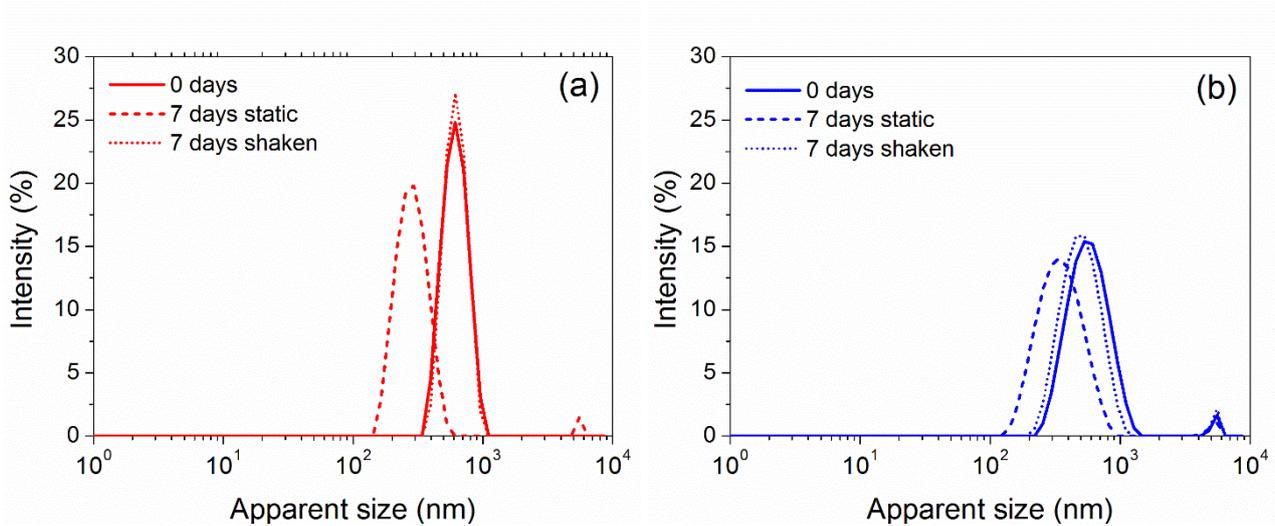

***Figure 4***: *Apparent size distribution for 0.025 wt% samples: (a) S-GnP/water nanofluid and (b) P-GnP/water nanofluid.*

Figure 5 shows the Z-average size, i.e. the intensity-weighted mean size obtained from a cumulants fit of the obtained autocorrelation function of intensity, as a function of the time after preparation.

The evolution of the Z-average size shows again the quasi-constant value obtained for the shaken samples, ~ 990 nm for S-GnP and ~ 540 nm for P-GnP. The static samples show again symptoms of sedimentation with decreases up to ~ 360 nm for S-GnP and ~ 330 nm for P-GnP after seven days. Furthermore, the lower decrease for the P-GnP nanofluid is evidenced and it should be noticed that the value obtained for this sample 24 h after preparation is similar to the initial one.

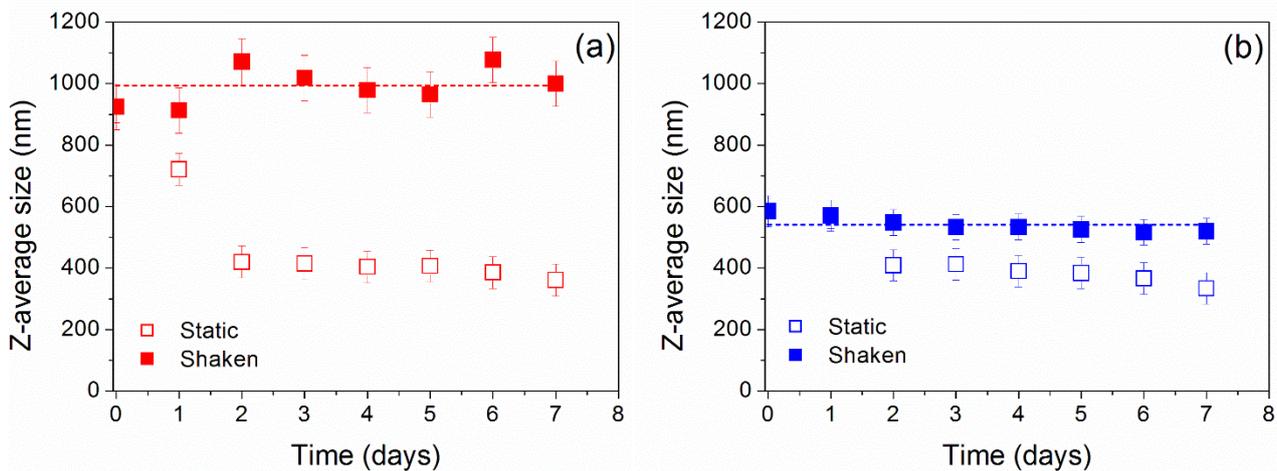

***Figure 5***: *Average Z-Size dependence on time after preparation for the 0.025 wt% samples. (a) S-GnP/water nanofluid and (b) P-GnP/water nanofluid.*

The complete stability analyses allow to conclude that moderately stable nanofluids were achieved with easily recoverable initial dispersion conditions. However, P-GnP/water nanofluids evidenced better long-term stability than S-GnP/water nanofluids.

### 3.2 Linear optical properties

Figures 6 and 7 show the experimental spectral extinction coefficient of P-GnP and S-GnP nanofluids, respectively. We can immediately observe that both nanoadditives show a higher extinction coefficient than water, especially in the UV, visible and near-infrared. Qualitatively speaking, both nanofluids show the typical UV plasmonic peak of the graphene-based nanostructures and an almost flat spectrum in the visible and near-infrared ranges [56]. The spectral features at around 1450 nm, 2000 nm and longer wavelengths are the water absorption bands; in particular the peak at around 2000 nm appears saturated, thus the height shown is not reliable. The enlargement of the 180-750 nm range in Fig. 6b allows to appreciate the spectrum of the sample at the lowest concentration. It is interesting to notice that P-GnPs extend their extinction tail in the whole investigated range, showing a well recognizable extinction coefficient increase even at the infrared local minima of water spectrum (around 1750 and 2250 nm). For the same values of nanopowder loading, the extinction coefficient is considerably higher in P-GnPs than in S-GnPs in the whole investigated wavelength range. Comparing both types of nanopowder, the shape of the UV plasmonic peak and of its tail appears different according to the different functionalization. Figure 8 shows the nanopowder influence to the extinction coefficient of samples decoupled from that of the base fluid, for the 0.025 wt% concentration. We can appreciate the higher and narrower plasmonic peak of P-GnPs and their higher plateau value in the visible and infrared. In the region of sunlight spectrum, which starts from 300 nm on, for the same concentration, P-GnPs show a nearly double extinction coefficient than S-GnPs, resulting in a better solar absorber. However also S-GnPs can achieve a comparable solar absorbance if a higher nanoparticle concentration will be taken into account.

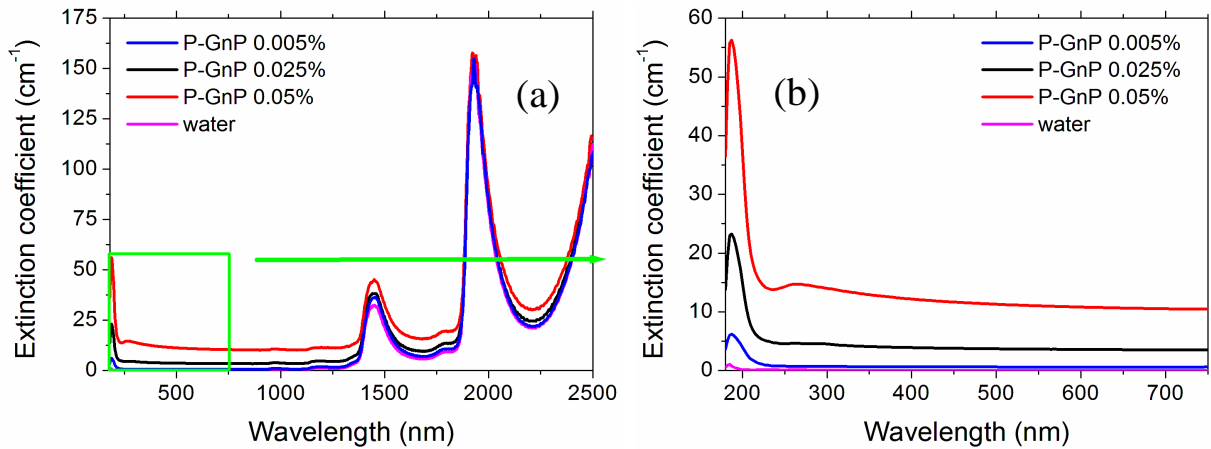

**Figure 6**: *Spectral extinction coefficient of P-GnP-based nanofluids with concentrations ranging from 0.005 wt% to 0.05 wt%. The water extinction coefficient is also reported for reference.*

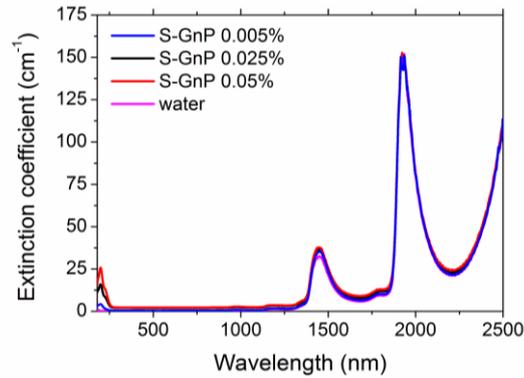

***Figure 7***: *Spectral extinction coefficients of S-GnP-based nanofluids at 0.025 wt% and 0.05 wt% concentrations. The water extinction coefficient is also reported for reference.*

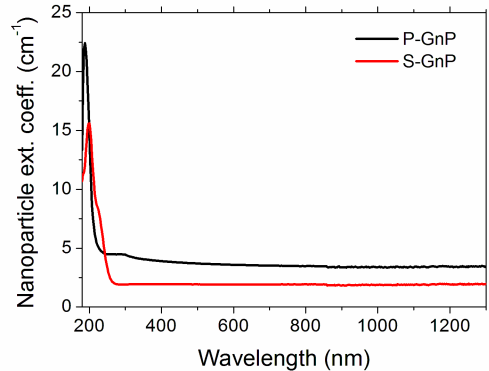

***Figure 8***: *Comparison of the nanoadditive contribution to the spectral extinction coefficients of nanofluids.*

The evidenced spectral properties result extremely important to characterize the sunlight extinction behavior of the nanofluids and their energy storage capability. The sunlight extinction fraction, EF, of the incident sunlight $I(\lambda)$ [57] which is extinct in the fluid after a propagation path of length x is given by the expression [20, 25]:

$$EF(x) = 1 - \frac{\int_{\lambda_{min}}^{\lambda_{MAX}} I(\lambda) \cdot e^{-\mu(\lambda)x} d\lambda}{\int_{\lambda_{min}}^{\lambda_{MAX}} I(\lambda) d\lambda} \qquad (1)$$

where $\mu(\lambda)$ is the spectral extinction coefficient and $\lambda_{min}$ and $\lambda_{MAX}$ are 300 and 2500 nm, respectively.

Figure 9 compares the calculated EF fraction for the investigated nanofluids at the concentrations of interest for solar absorber applications (0.025 wt% and 0.05 wt%) to that of the base fluid. We can see how the addition of the different kinds of functionalized graphene nanoplatelets dramatically modifies the interaction of the fluid with solar radiation, producing an almost complete sunlight extinction after a path length, which, depending on the nanoadditive type and concentration, ranges from 5 to 30 mm. This modification indicates that the light extinction after a path as long as 50 mm in the nanofluids is about 250% higher than that of pure water. As expected from the small differences in the extinction spectra, doubling the concentration in S-GnPs decreased the length for 99% sunlight extinction only by about 3 mm (from 22 to 19 mm), while the concentration doubling in P-GnPs more heavily affected the extinction length (99% extinction length shortened from about 12 to 4 mm), in agreement with the largest differences in the spectral extinction coefficient shown by these nanoadditives (Figure 6).

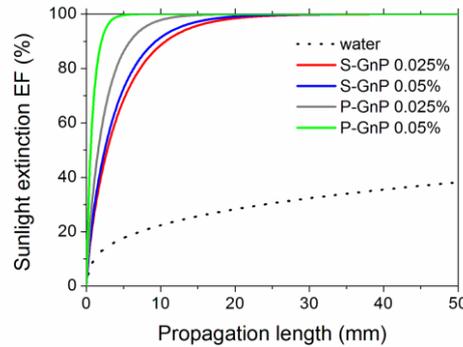

***Figure 9****: Calculated sunlight extinction as a function of the propagation length within the nanofluids.*

It has to be considered that the type and concentration of nanoadditives also affect the spatial distribution S(x) of the stored energy inside the nanofluid volume. This distribution, for a cold fluid in absence of convective mixing is given by the expression:

$$S(x) = \frac{\int_{\lambda_{min}}^{\lambda_{MAX}} I(\lambda) \cdot \mu(\lambda) \cdot e^{-\mu(\lambda)x} d\lambda}{\int_{\lambda_{min}}^{\lambda_{MAX}} I(\lambda) d\lambda} \qquad (2)$$

Plots of the calculated stored power distributions along the light propagation direction are reported in Figure 10. Distributions refer to a single-sided irradiated nanofluid as in the case of a generic radial direction in transverse section of the absorber tube in a sunlight parabolic-linear collector, or in the case of the thickness direction in a flat-plate collector. The curves in Fig. 10 have been respectively normalized referring to the highest value of the distribution for the 0.05 wt% P-GnP sample, thus they have the meaning of relative values. As we can see (Fig. 10), the higher are extinction coefficients, the more localized is sunlight extinction in the first layers of fluid. For the highest concentration P-GnP sample, the energy is stored in a very small depth near the surface and inner layers are not directly heated by the light at all, resulting in a strong distribution gradient. As the nanoadditive concentration decreases, the stored energy distribution penetrates more deeply in the sample, producing, after the first steep gradient, a uniform profile in the inner layers. For S-GnPs samples, sunlight penetrates even more deeply in the volume, and further internal layers can be heated. As apparent from Fig. 10, the energy distribution for pure water is lower than those of all nanofluid samples, even for the less-concentrated S-GnPs nanofluid. As a final comment of this section, it is important to notice that both the sunlight extinction and the stored energy distribution can be tailored and adapted to the system geometry by tuning the extinction coefficient. That is possible adjusting the concentration and/or the type of nanoparticles. In fact, the literature reports, for solar collectors allowed to operate under optimal conditions both as for the nanofluid and the system architecture design, efficiency increases from 5 to 10% with respect to surface solar absorbers [58-60].

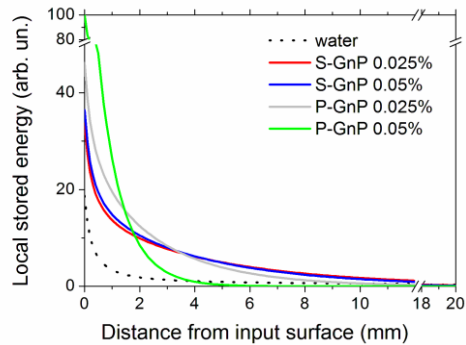

***Figure 10***: *Stored energy distribution in the nanofluids as a function of the distance from the sunlight input surface.*

### 3.3 Nonlinear optical properties

Due to the high extinction coefficient of the samples with 0.05 wt% nanoadditive concentration, preventing any transmitted signal to be recorded at the output after the 10-mm length cell, high intensity experiments were carried out on the lower 0.005 wt% and 0.025 wt% concentrations.

Figures 11 and 12 show the wavelength-dependent output energy measurements for the two investigated concentrations for both nanofluid types. All samples show a marked nonlinear behaviour. At 355 and 532 nm, the samples with the lowest concentration show a classical optical limiting behaviour characterized by no sign inversion in the concavity of the curve deviating from linearity and by an asymptotic-like trend, which is more recognizable, within the investigated energy range, in P-GnP-based suspensions. At 1064 nm, where the laser is able to supply the highest energy, the larger energy range experimentally accessible reveals a concavity change which, in carbon-nanohorn based suspensions, was explained in terms of laser-induced sample damage [61]. In all cases, the P-GnP samples show both the lowest starting linear transmittance and the most pronounced nonlinearity.

The higher 0.025 wt% concentration shows, for both samples and at all wavelengths, a more complex nonlinear behaviour, with a non-monotonic trend as a function of the input energy, characterized by a decrease after a local maximum, and, for the sample S-GnPs at 1064 nm, even by a further increase for input energy higher than 10 mJ. Moreover, the two samples show very large difference in transmitted energy, so that two different ordinate axes are used.

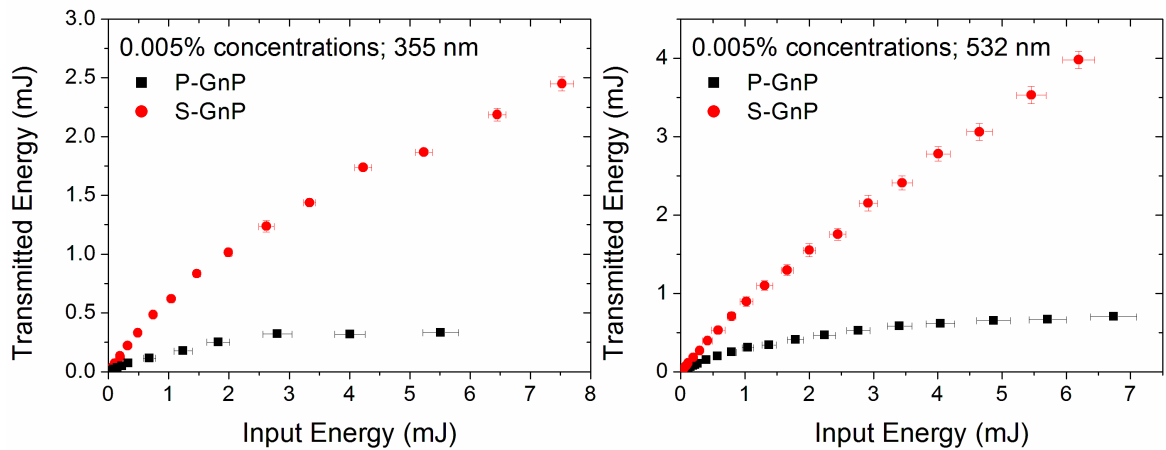

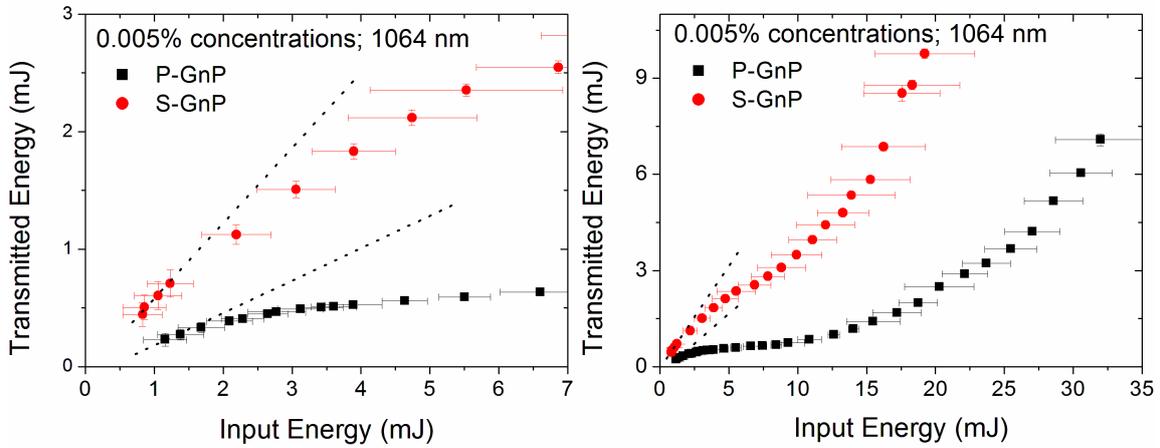

**Figure 11.** *Transmitted energy as a function of the incident energy under different wavelengths and for the 0.005 wt% nanoadditive concentrations. The lower plots are both referred to the 1064 nm wavelength (the dotted lines are a visual reference of the linear regime).*

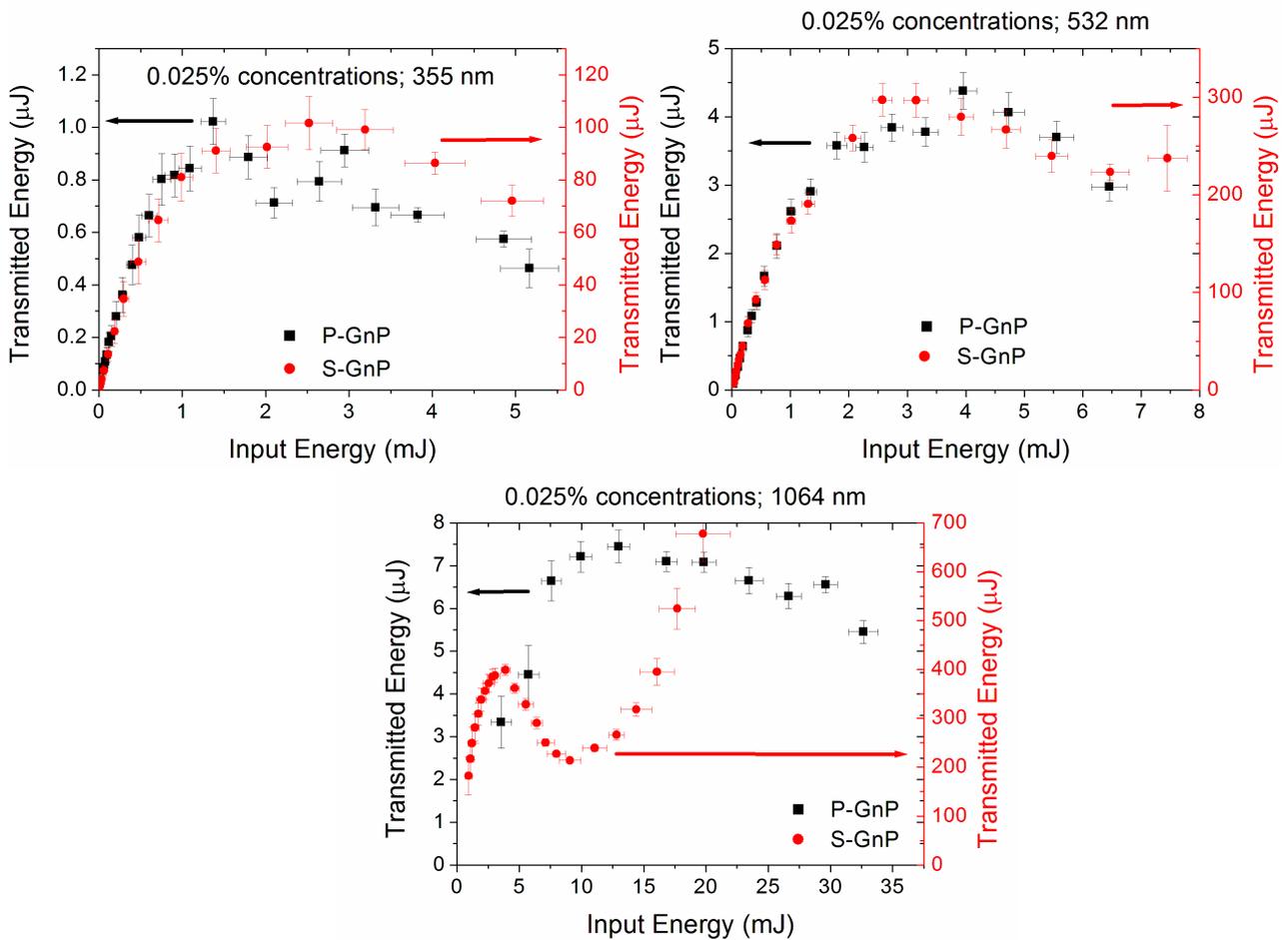

**Figure 12.** *Transmitted energy as a function of the incident energy under different wavelengths and for the 0.025 wt% nanoadditive concentrations.*

Figure 13 compares, for fixed nanoparticle type and concentration, the effect of changing the laser wavelength. At all wavelengths and concentrations, the energy transmitted by P-GnP samples is considerably lower than that of the corresponding S-GnP sample, in agreement with the lower linear extinction coefficient, and, as noticed above, the nonlinearity is more pronounced. For fixed nanoadditive type, the wavelength-dependent nonlinear behavior changes with the concentration. In fact, even if the minimum transmittance is always obtained at 355 nm, the relative hierarchy of green and infrared wavelength changes with the concentration. This seems to suggest different nonlinear mechanisms acting at the lowest and highest concentration, at least for 532 nm and 1064 nm wavelength. Moreover, for fixed sample and concentration, it is interesting to notice that the minimum detectable output power is higher at 1064 nm than at 532 nm, despite that the linear extinction coefficient at these wavelengths shows similar values, again suggesting a stronger nonlinearity, or, at least, a lower nonlinearity energy threshold at 1064 nm.

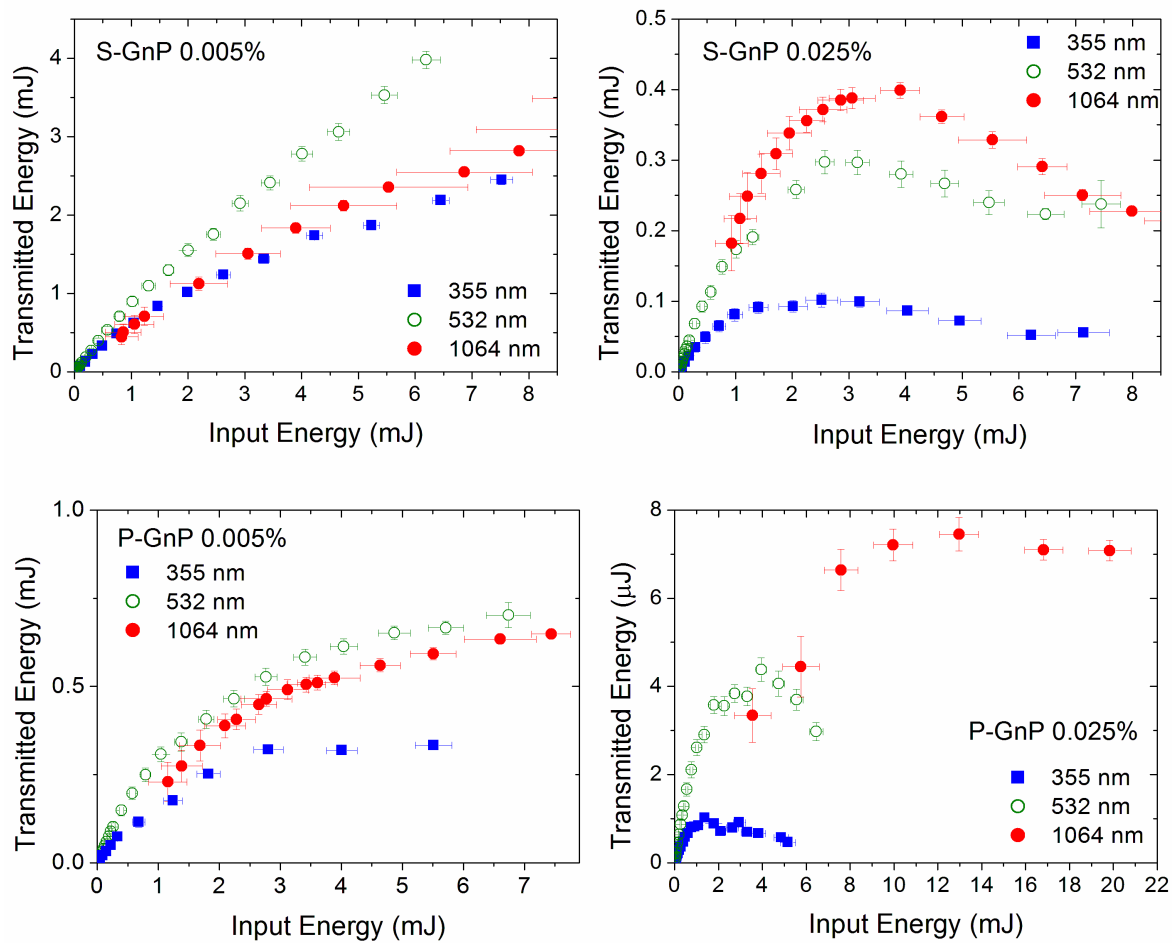

**Figure 13.** *Transmitted energy as a function of the incident energy for fixed nanoadditive type and concentration.*

The phenomenon of optical limiting in colloids can be explained by two mechanisms: nonlinear absorption or nonlinear scattering [62], depending on the nature of nanoparticles and on the properties of the suspending fluids [63]. The effect actually occurring can be identified applying semi-empirical models. Nonlinear absorption can be due to different effects like reverse saturable absorption, two-photon absorption etc [63]. It can modeled introducing a nonlinear term in the Lambert-Beer law [64]:

$$\frac{dF(z)}{dz} = -\mu F - \frac{\mu \sigma_1}{2\hbar\omega} F^2 \tag{3}$$

where F(z) is the energy fluence at the z position in the sample, $\mu$ is the linear extinction coefficient, $\sigma_1$ is the Excited State Absorption (ESA) cross section. If we solve this equation following the method described in [65, 66], we can link the output energy from the nanofluid, $E_{out}$, to incident energy $E_{in}$, obtaining the following relationship:

$$E_{out} = \frac{T_s^2 e^{-\mu L} E_{in}}{1 + 0.1 \, E_{in}/E_t} \tag{4}$$

Where $T_s^2$ is a coefficient describing the reflection losses on the walls of the cuvette and L is the thickness of the sample. The term $e^{-\mu L}$ describes the linear extinction, and $E_t$ is the threshold energy defined as the energy where the transmittance decreases to the 90% of its original value [64]. This model will be identified as LB in the following.

On the other hand, a model where optical limiting is only due to nonlinear refraction has been developed in [67]. In the cited reference, nonlinear properties are ascribed to nonlinear scattering between laser photons and vapor bubbles yielded by the vaporization of the base fluid surrounding the nanoparticle and/or by ionization and vaporization of the nanoparticles:

$$E_{out} = T_s^2 E_t \, e^{-\mu L} \left( 1 + \frac{\mu^e}{\mu} \left( \frac{E_{in}}{E_t} - 1 \right) \right)^{\frac{\mu}{\mu^e}} \tag{5}$$

where $\mu^e$ is the nonlinear extinction coefficient and $E_t$ is the energy threshold for bubble formation, both obtained from a fitting procedure of experimental data. This model is labelled as MMJ in the following.

A series of numerical tests were carried out to verify whether theoretical models are compatible with our experimental data, and, in the positive case, to estimate the values of the model parameters. The procedure consisted in a standard least-squares approach, where the objective function is the sum S of the squared differences between the observed $E_{out}$ values and the corresponding computed values for each $E_{in}$, as described in Ref. [61].

None of the considered models can reproduce non-monotonic nonlinear trends, which require thus further investigation. The cases when the fitting was successful are shown in Figure 14. We can see that, except the case 0.005 wt% S-GnP and 532 nm, the only acceptable model is the nonlinear Lambert-Beer (Eq. 4). Table 2 lists the obtained energy threshold values. If we compare the values obtained from the nonlinear Lambert-Beer model, we can see that the threshold increases with the wavelength and that it is always lower in P-GnPs than in S-GnPs samples. In the only case when fitting by the MMJ model was acceptable, we can see that the MMJ threshold value is one order of magnitude lower than the Lambert-Beer one. This result is not surprising, because the two models account for different physical phenomena (nonlinear absorption and nonlinear refraction).

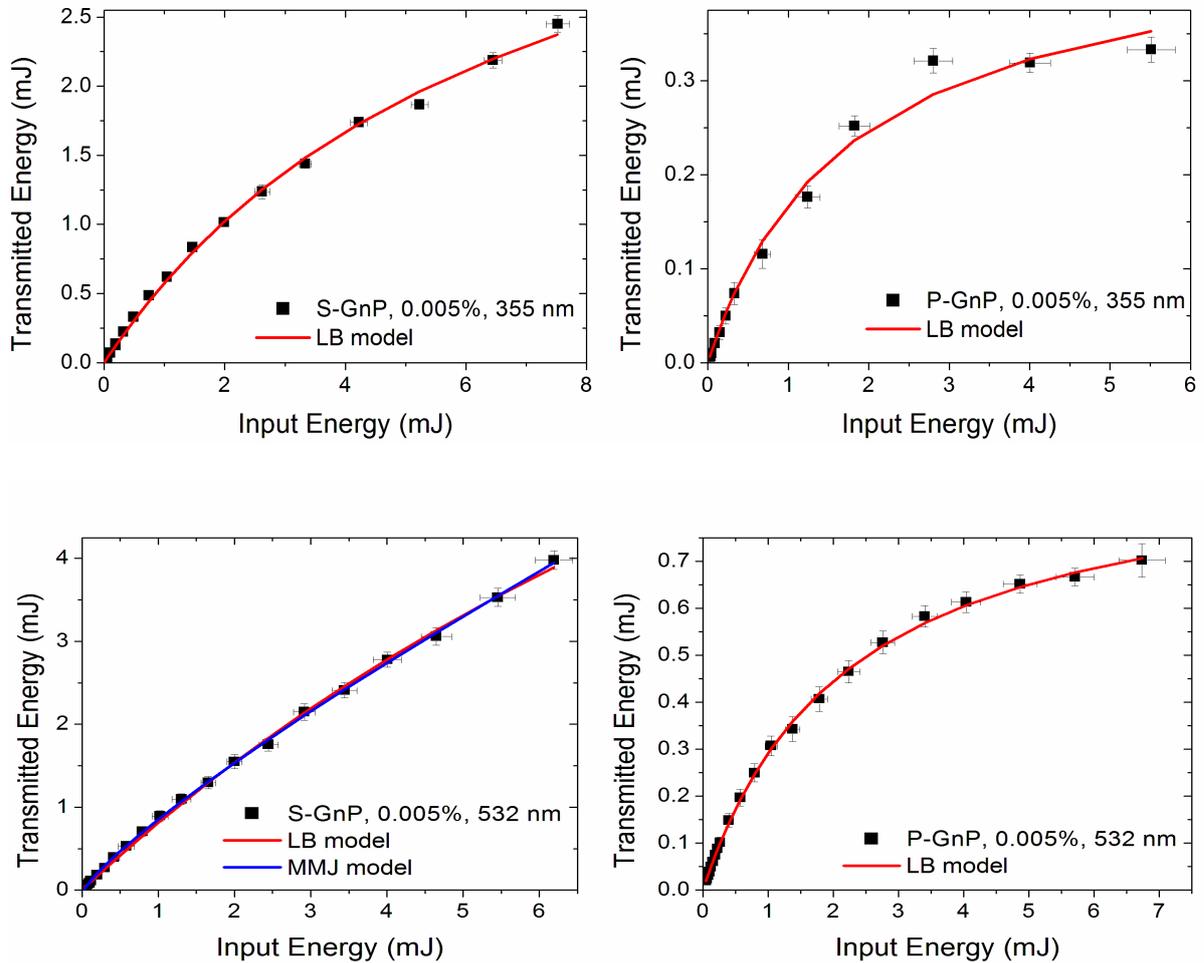

***Figure 14.** Fitting of experimental data with Eqs. 4 and 5.*

| Sample, concentration | Wavelength (nm) | Energy threshold (mJ), model |
|---|---|---|
| 0.005 wt%, P-GnPs | 355 | $0.18 \pm 0.02$, LB |
| | 532 | $0.22 \pm 0.02$, LB |
| 0.005 wt%, S-GnPs | 355 | $0.70 \pm 0.08$, LB |
| | 532 | $1.7 \pm 0.5$, LB |
| | | $0.17 \pm 0.02$, MMJ |

***Table 2.** Energy threshold values obtained from the semi-empirical model fitting.*

## 4. Conclusions

Two different nanosuspensions of polycarboxylate chemically modified graphene nanoplatelets (P-GnP) and Sulfonic acid-functionalized graphene nanoplatelets (S-GnP) dispersed in water were tested, at three different concentrations (0.005 wt%, 0.025%wt and 0.05 wt%). Dispersions, and especially P-GnP, showed a satisfying long-term stability. Spectrophotometric measurements allowed to obtain the spectral extinction coefficients and to evaluate the potential for direct solar absorber applications. For both colloid types, the sunlight radiation is nearly completely extinguished in 5-20 mm path length, with the differences among the samples connected to the concentration and to the different functionalization of nanoadditives. The nonlinear optical measurements, carried out on 0.005 wt% and 0.025% wt concentrations, evidenced a non-linear behaviour for both nanoadditive types, both concentrations, and at all the three test wavelengths (ultra-violet, visible, and near-infrared), demonstrating appealing broadband characteristics. Fitting of nonlinear data with semi-empirical models showed thresholds between 0.18 and 1.7 mJ, and the prevalence of nonlinear absorption effects for the lowest concentrations. More marked and non-monotonic nonlinear behaviours were obtained at the highest concentrations, seeming to suggest the occurrence of concurrent nonlinear mechanisms, likely involving the production of bubbles. The energy densities reached in the experiments are compatible with solar concentration systems [68], thus opening interesting perspectives for the further application of these nanofluids in solar vapor generation and solar desalination [69].

### Acknowledgments

This work was partially supported by EU COST Action CA15119: Overcoming Barriers to Nanofluids Market Uptake (NANOUPTAKE) in the framework of the Short Term Scientific Mission program. This work was also partially supported by "Ministerio de Economía y Competitividad" (Spain) and FEDER program through ENE2014-55489- C2-2-R and ENE2017-86425-C2-1-R projects. J. P. V. acknowledges FPI Program of "Ministerio de Economía y Competitividad". Thanks are due to Mr. M. D'Uva and Mr. M. Pucci (INO-CNR) for technical assistance.